\documentclass[journal]{IEEEtran}

\usepackage{ulem}
\usepackage{soul,color}
\usepackage{srcltx}
\usepackage{multirow}
\usepackage[tbtags,sumlimits,nointlimits,reqno]{amsmath}
\allowdisplaybreaks[4]
\usepackage{stfloats}
\usepackage{subfigure}
\usepackage{psfrag}
\usepackage{color}
\usepackage{amssymb}
\usepackage{cite}
\usepackage{graphicx}
\ifCLASSINFOpdf
\else
\fi
\usepackage{amsmath}
\usepackage{threeparttable}

\begin{document}

\title{Spoof Surface Plasmon Polariton Leaky-Wave Antennas using Periodically Loaded Patches above PEC and AMC Ground Planes}

\author{QingLe~Zhang,~\IEEEmembership{Student~Member,~IEEE,}
        Qingfeng~Zhang,~\IEEEmembership{Senior~Member,~IEEE,}
        and~Yifan~Chen,~\IEEEmembership{Senior~Member,~IEEE}

\thanks{Manuscript received April 2017.}
\thanks{This work is measured at the State Key Laboratory of Millimeter Waves, Partner Laboratory in City University of Hong Kong, Kowloon, Hong Kong.}
\thanks{Q. Zhang and Q.L. Zhang are with the Department of Electronics and Electrical Engineering, Southern University of Science and Technology, Shenzhen, Guangdong, China (e-mail: zhang.qf@sustc.edu.cn).}

\thanks{Y. Chen is with University of Waikato, Hamilton, New Zealand.}
}


\maketitle

\begin{abstract}
This paper proposes two spoof surface plasmon polariton (SSPP) leaky-wave antennas using periodically loaded patches above perfect electric conductor (PEC) and artificial magnetic conductor (AMC) ground planes, respectively. The SSPP leaky-wave antenna is based on a SSPP transmission line, along which circular patches are periodically loaded on both sides to provide an additional momentum for phase matching with the radiated waves in the air. The PEC and AMC ground planes underneath the antenna reflect the radiated waves into the upward space, leading to an enhanced radiation gain. Both PEC- and AMC-grounded antenna prototypes are fabricated and measured in comparison with the one without any ground plane. The experimental results show that the PEC and AMC ground planes increase the radiation gain by approximately 3 dB within the operational frequency range 4.5-6.5 GHz. It also demonstrates that the AMC-grounded leaky-wave antenna, with a thickness of 0.08${\lambda _0}$ at 6 GHz, features more compact profile than the PEC-grounded one (with a thickness of 0.3${\lambda _0}$ at 6 GHz).
\end{abstract}


\IEEEpeerreviewmaketitle

\section{Introduction}

\IEEEPARstart{M}{icrostrip} antennas have been widely used in wireless communication systems due to low profile, light weight, low fabrication cost and easy integration with other RF devices. However, the microstrip line has huge ohmic loss and unwanted radiation loss at a high frequency~\cite{1Garg2001}. The microstrip antennas suffer from low efficiency as well because the radiator gets too close to the ground plane. All these lead to a big challenge for microstrip antennas to be applicable at millimeter-wave (MMW) and terahertz (THz) frequencies.

Spoof surface plasmon polariton (SSPP) transmission line, inspired by surface plasmon polaritons at optics, has attracted much attention recently due to its highly confined field, subwavelength operational resolution, and relatively low ohmic loss at MMW and THz frequencies. It has inspired various interesting research works, such as wave guiding structures~\cite{2Liao2014, 3Kianinejad2016,4Yuan2016,5Kianinejad2015}, filters\cite{6Zhao2016,7Qian2016,8Xu2016}, wave splitters \cite{9Jin2014}, and antennas \cite{10Zheng2016,11Yin2015, 12Yin2016, 13Xu2015, 14Gu2016, 15Kianinejad2016, 16Kianinejad2017, 17Yin2016 }. Basically, SSPP transmission line supports a slow transversal magnetic (TM) wave, and hence does not radiate due to lack of enough momentum. Inspired by phase-gradient metasurface, Xu \emph{et al.} periodically modulate the SSPP metallic strips to add an additional momentum to the propagating waves and hence allow them to radiate into the air~\cite{13Xu2015}. Gu \emph{et al.} proposed a leaky-wave antenna using non-uniformly modulated plasmonic waveguide~\cite{14Gu2016}, and employed an asymmetrical profile to get a high-efficiency radiation at the broadside. In~\cite{16Kianinejad2017}, it reported a single-layer leaky-wave antenna without loading terminations. In \cite{17Yin2016}, a circular patch array fed by planar SSPP transmission line was presented.

Since SSPP is a single-conductor line without any ground plane, most SSPP leaky-wave antennas exhibit omnidirectional H-plane radiations due to the quasi-circular symmetrical profiles. This is not preferred in some applications that require directional beams. In this paper, we propose a SSPP leaky-wave antenna using periodically loaded circular patches incorporating with either a perfect electric conductor (PEC) or an artificial magnetic conductor (AMC) ground plane underneath the antenna to achieve directional beams in the H plane. It shows that both ground planes improve the radiation gain by approximately 3 dB. The optimum distance of the PEC ground plane is  $0.3\lambda _0$ at 6 GHz. In contrast, the AMC ground plane gets much closer to the radiator, around $0.08\lambda _0$ at 6GHz. Therefore, the AMC-grounded SSPP leaky-wave antenna performs better in low-profile applications.




\section{SSPP Transmission Line}

The configuration of the SSPP transmission line is shown in Fig. 1(a). It is fabricated on an 1-mm-thick F4B substrate (with $\epsilon_r=2.65$ and $\tan\sigma=0.003$). Since SSPP transmission line is a single-conductor line not compatible with the SMA connector, a smooth transition into $50~\Omega$ coplanar waveguide (CPW) is designed for both impedance transformation and mode conversion. The overall width and length of the SSPP transmission line including the CPW transition are $w_0=70$~mm and $L_0=328$~mm, respectively. For the CPW transmission line at the ports, the strip width, gap size, and length are chosen as $w_c=2.3$~mm, $g_c=0.2$~mm, and $L_c=10$~mm, respectively. To accommodate the single-conductor SSPP line, the CPW ground planes gradually fade away in an exponential function, $y = f(x) = {e^{\alpha {\rm{x}}}} - 1$
where $\alpha = \ln(w_0/2-w_c/2-{g_c}+1)/{L_1}$ and $L_1=47.5$~mm. The gradually tapered ground plane reduces the equivalent capacitance and hence increases the characteristics impedance for matching with the SSPP line. In fact, SSPP transmission line can also be regarded as a multi-conductor line with ground planes placed at an infinite distance. Thus, this tapered ground plane is a natural transition for SSPP line. Besides, the exponentially tapered ground plane also produces a longitudinal electric field, which provides a smooth TEM-to-TM mode conversion.

\begin{figure}
\centering
\psfrag{a}[c][c]{\footnotesize (a)}
\psfrag{b}[c][c]{\footnotesize (b)}
\psfrag{z}[c][c]{\footnotesize (c)}
\psfrag{m}[c][c]{\footnotesize $\beta p$}
\psfrag{n}[c][c]{\footnotesize Frequency (GHz)}
\psfrag{q}[r][c]{\footnotesize Air Line}
\psfrag{u}[c][c]{\footnotesize SSPP}
\includegraphics [width=7.5cm]{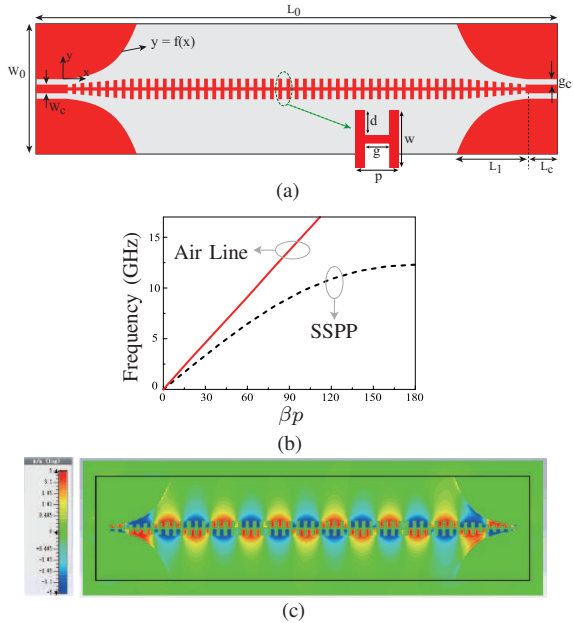}
\caption{(a) The geometry configuration of the SSPP transmission line; (b) Dispersion curve of the SSPP unit cell; (c) Calculated magnetic field $H_z$ distribution at 6 GHz.}
\end{figure}

The SSPP transmission line employs an H-shape corrugated strip, whose parameters are $p=5.5$~mm, $w=9.5$~mm, $d=4$~mm and $g=3$~mm.
 This H-shpae unit cell is analysed by the Eigen-mode solver of CST Microwave Studio, and the computed dispersion curve is shown Fig. 1(b). Note that, the supported mode is always below the air line [shown as the red solid line in Fig. 1(b)], and hence it is a bounded slow-wave mode. The dispersion curve gradually bends away from the air line when the operational frequency increases up to the cutoff frequency at $12.3$~GHz. Since the momentum of this mode is always smaller than that of the wave in the air, the field does not radiate into the air and hence is well confined along the line. Fig. 1(c) displays the computed magnetic field on the substrate at 6 GHz. One clearly observes a well confined surface mode propagating along the SSPP transmission line. Also, the field is alternating between the two sides of the H-shape line. This allows one to load patches on both sides of the SSPP line to achieve leaky-wave radiations, which will be introduced in the forthcoming section.

\section{SSPP Leaky-Wave Antenna}

 As illustrated in Figs. 1(b) and (c), SSPP transmission line supports a slow-wave surface mode, which does not radiate due to lack of enough wave momentum. To achieve radiation, it requires an additional momentum, which may be realized either by periodically modulating the profile~\cite{13Xu2015}, or by periodically loading circular patches along the SSPP line~\cite{17Yin2016}. Here, we employ patch-loading on both sides of the SSPP line, as shown in Fig. 2(a). The circular patches are placed at the positions with maximum magnetic field in Fig. 1(c). Since the maximum points are alternating on two sides of the line, the patches are loaded in an alternating fashion as well. The two patches in one period are excited in phase since the magnetic fields at the two points have identical directions. According to the dispersion curve in Fig. 1(b), the SSPP guided wavelength is ${\lambda _g} = 2\pi /\beta  = 36$  mm at 6 GHz. In order to achieve a broadside radiation at 6 GHz, the patch loading period should be the same as the guided wavelength, i.e. $a=\lambda _g=36$~mm. The patch radius, $R_1$, and the distance from the SSPP line, $b/2$, are used to control the coupling level and hence the radiated power of each element. The optimized parameters are $R_1=\lambda _g/4=9$~mm and $b=28.5$~mm. Fig.~2(b) shows the magnetic field $H_z$ distribution of the patch-loaded SSPP leaky-wave antenna at 5.6~GHz. The wave power is continuously coupled into the circular patches when it propagates along the SSPP transmission line. Note that, all the patches on both sides of the SSPP line are excited almost in phase, which leads to a broadside radiation. The broadside frequency (5.6~GHz) is slightly different from the unloaded case (6.0~GHz) due to the loading effect brought by circular patches. Compared with the single-side loading structure in~\cite{17Yin2016}, this patch-loading on both sides of the line greatly enhances the efficiency of space utilization.

\begin{figure}
\centering
\includegraphics [width=7.5cm]{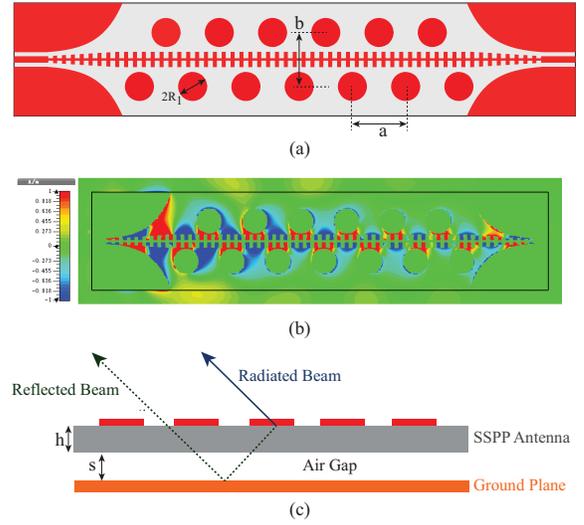}
\caption{(a) Configuration and (b) magnetic field $H_z$ distribution (at 5.6~GHz) of the patch-loaded SSPP leaky-wave antenna; (c) Ground plane underneath the SSPP leaky-wave antenna.}
\end{figure}

In addition to patch-loading on both sides of the SSPP line, the proposed leaky-wave antenna also employs a ground plane underneath it, as shown in Fig. 2(c). The ground plane reflects the waves into the upper space, and hence increases the radiation gain. We explore both PEC and AMC ground planes in this research. The distance between the ground plane and the antenna, $s$, should be optimized for different cases. The ideal distance for a PEC ground plane should be around a quarter of the space wavelength so that the reflected wave are in phase with the directly radiated wave. In this design, the optimum distance for the PEC ground plane is $s=0.3{\lambda _0}=15$~mm, where $\lambda_0$ is the space wavelength at 6 GHz. In contrast, the AMC ground plane can be placed much closer to the antenna. Fig. 3(a) depicts the geometry of the AMC ground, which is formed by periodic square patches printed on an F4B substrate with an air-gapped metal plane underneath it. The thickness of the substrate $h_1$, the air gap $h_2$, the patch size $p_2$, and the patch period $p_1$, are optimized to maximize the operational bandwidth. The optimum parameters are $h_1=2$~mm, $h_2=2$~mm, $p_2=8.3$~mm, and $p_1=10$~mm. Fig. 3(b) shows the reflection phase under perpendicular incidence. Note that the operational frequency range is $4.45-7.04$~GHz, within which the reflection phase stays between $+90^{\circ}$ and $-90^{\circ}$. When placing this AMC underneath the SSPP leaky-wave antenna, the optimum distance, $s=4$~mm, is only $26.7\%$ of the PEC case. Even counting in the AMC thickness, the overall distance ($h_1+h_2+s=8$~mm) is still $53\%$ of the PEC case. Therefore, AMC-grounded SSPP leaky-wave antenna features more compact profile than the PEC-grounded one.

\begin{figure}
\centering
\psfrag{a}[c][c]{\footnotesize (a)}
\psfrag{b}[c][c]{\footnotesize (b)}
\psfrag{f}[c][c]{\footnotesize Frequency (GHz)}
\psfrag{g}[c][c]{\footnotesize Reflection Phase (Deg)}
\psfrag{s}[c][c]{\footnotesize Substrate}
\psfrag{m}[c][c]{\footnotesize Metal}
\psfrag{r}[c][c]{\footnotesize Air}
\psfrag{c}[c][c]{\footnotesize $p_1$}
\psfrag{d}[c][c]{\footnotesize $p_2$}
\psfrag{e}[c][c]{\footnotesize $h_1$}
\psfrag{h}[l][c]{\footnotesize $h_2$}
\includegraphics [width=8.6cm]{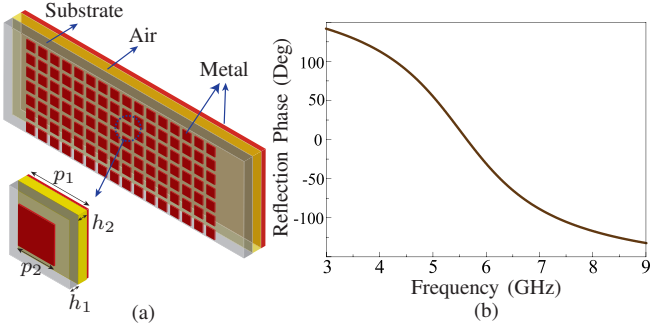}
\caption{(a) Configuration of the AMC ground plane; (b) Reflection phase of the AMC plane under perpendicular incidence.}
\end{figure}

\section{Experimental Validation}

To experimentally verify the performance of the PEC-grounded and AMC-grounded SSPP leaky-wave antennas, they are fabricated and measured in comparison with the one without any ground plane. In Fig. 4(a), the top figure shows the fabricated prototype of the SSPP leaky-wave antenna with loaded circular patches on both sides of the SSPP line, and the bottom figure displays the fabricated prototype of the AMC ground (in top view). Since the PEC ground is purely metals, its fabricated prototype is not shown here. Fig. 4(b) illustrates the assembly photograph of both PEC-grounded and AMC-grounded SSPP leaky-wave antennas. Different layers are piled up with finely tuned air gaps. Plastic screws are placed at four edge corners to fix the distance between layers. One clearly observes that AMC-grounded SSPP leaky-wave antenna is much thinner than the PEC-grounded one. All the fabricated prototypes are measured in an Agilent vector network analyzer E5071C for reflection responses and an Anechoic Chamber for radiation patterns and gains.

\begin{figure}
\centering
\includegraphics [width=8.6cm]{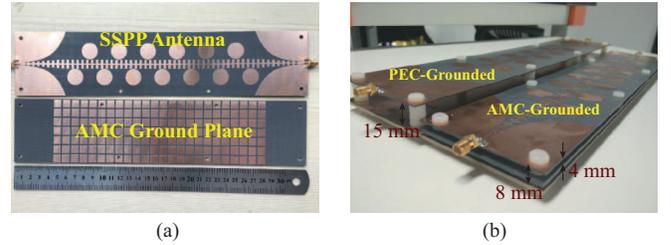}
\caption{Photograph of the fabricated prototypes: (a) SSPP leaky-wave antenna and AMC ground plane, (b) Comparison of PEC-grounded antenna and AMC-grounded antenna.}
\end{figure}

\begin{figure*}
\centering
\psfrag{a}[c][c]{\footnotesize (a)}
\psfrag{b}[c][c]{\footnotesize (b)}
\psfrag{c}[c][c]{\footnotesize (c)}
\psfrag{f}[c][c]{\footnotesize Frequency (GHz)}
\psfrag{t}[c][c]{\footnotesize Theta (Deg)}
\psfrag{r}[c][c]{\footnotesize Radiation Pattern (dB)}
\psfrag{g}[c][c]{\footnotesize Radiation Gain (dBi)}
\psfrag{s}[c][c]{\footnotesize Reflection (dB)}
\psfrag{x}[l][c]{\footnotesize Sim-w/o-PEC}
\psfrag{y}[l][c]{\footnotesize Sim-with-PEC}
\psfrag{z}[l][c]{\footnotesize Mea-w/o-PEC}
\psfrag{d}[l][c]{\footnotesize Mea-with-PEC}
\psfrag{m}[l][c]{\footnotesize Measured}
\psfrag{n}[l][c]{\footnotesize Simulated}
\includegraphics [width=18cm]{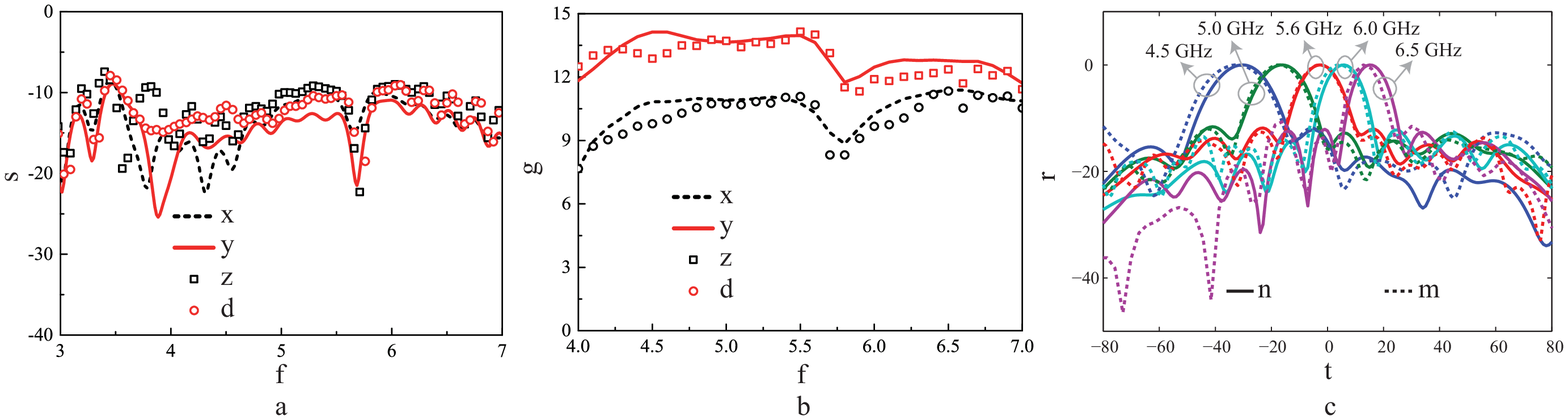}
\caption{Simulated and measured (a) reflection responses and (b) radiation gain of the SSPP antenna with and without PEC grounds; (c) Simulated and measured radiation patterns of PEC-grounded SSPP antenna at different frequencies.}
\end{figure*}

\begin{figure*}
\centering
\psfrag{a}[c][c]{\footnotesize (a)}
\psfrag{b}[c][c]{\footnotesize (b)}
\psfrag{c}[c][c]{\footnotesize (c)}
\psfrag{f}[c][c]{\footnotesize Frequency (GHz)}
\psfrag{t}[c][c]{\footnotesize Theta (Deg)}
\psfrag{r}[c][c]{\footnotesize Radiation Pattern (dB)}
\psfrag{g}[c][c]{\footnotesize Radiation Gain (dBi)}
\psfrag{s}[c][c]{\footnotesize Reflection (dB)}
\psfrag{x}[l][c]{\footnotesize Sim-w/o-AMC}
\psfrag{y}[l][c]{\footnotesize Sim-with-AMC}
\psfrag{z}[l][c]{\footnotesize Mea-w/o-AMC}
\psfrag{d}[l][c]{\footnotesize Mea-with-AMC}
\psfrag{m}[l][c]{\footnotesize Measured}
\psfrag{n}[l][c]{\footnotesize Simulated}
\includegraphics [width=18cm]{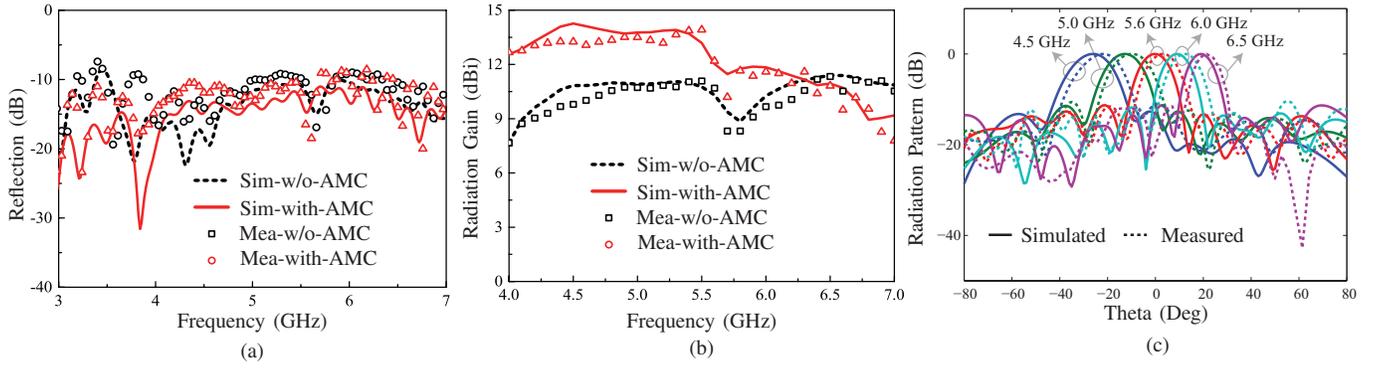}
\caption{Simulated and measured (a) reflection responses and (b) radiation gain of the SSPP antenna with and without AMC grounds; (c) Simulated and measured radiation patterns of AMC-grounded SSPP antenna at different frequencies.}
\end{figure*}

One firstly compares the SSPP leaky-wave antennas with and without a PEC ground plane. Fig. 5(a) shows the simulated and measured reflection responses of both cases. The measured responses agree well with the simulated ones within the whole measured frequency range from 3 GHz to 7 GHz. Also, the reflections of both SSPP leaky-wave antennas with and without PEC ground plane are below $-10$~dB within the frequency band 4-7 GHz (with a relative bandwidth of $54.5\%$). It turns out that the PEC ground plane does not affect the impedance matching of the SSPP leaky-wave antenna. This is possibly attributed to the highly confined field of SSPP transmission line and its relatively large separation (around $0.3\lambda_0$) from the PEC ground. Although the PEC ground plane does not affect the reflection response, it increases the radiation gain by reflecting the downward waves into the upper space. Fig.~5(b) compares the radiation gains of SSPP leaky-wave antennas with and without a PEC ground plane. The PEC ground plane increases the radiation gain by almost 3 dB within the frequency range 4-5.5 GHz, inspite of a smaller increment after 5.5 GHz. Furthermore, the measured responses agree well with the simulated ones for both cases. The maximum measured radiation gain is around 14.1~dBi for the PEC-grounded SSPP leaky-wave antenna. Fig. 5(c) shows the simulated and measured $xz$-plane radiation patterns of the PEC-grounded antenna at different frequencies. The measured beam, in a good agreement with the simulated one, continuously scans from $-32^{\circ}$ to $13^{\circ}$ as the frequency increases from 4.5~GHz to 6.5~GHz.

For the AMC-grounded SSPP leak-wave antenna, its simulated and measured reflection responses are shown in Fig. 6(a). Similar to the PEC-grounded case, the measured reflection response, in good agreement with the simulated one, is below $-10$~dB within the frequency band 4-7 GHz. Therefore, the AMC ground does not affect the impedance matching as well although it stays much closer to the SSPP leaky-wave antenna. Fig. 6(b) compares the radiation gains of the SSPP leaky-wave antennas with and without an AMC ground plane. The AMC ground plane also increases the radiation gain by almost 3 dB within the frequency range 4-5.5 GHz, inspite of a degradation after 5.5 GHz. The maximum measured gain is around 13.9 dBi. Fig. 6(c) shows the simulated and measured $xz$-plane radiation patterns of the AMC-grounded antenna at different frequencies. The measured beam, continuously scanning from $-22^{\circ}$ to $21^{\circ}$ as the frequency increasing from 4.5~GHz to 6.5~GHz, exhibit a constant shift compared with the simulated result. This slight shift is possibly attributed to the tolerances mainly contributed by the two air gaps between layers, which is brought by manual assembly.

\section{Conclusion}

SSPP leaky-wave antennas using loaded patches above a PEC and an AMC ground plane, respectively, have been introduced. The results reveal that both ground planes increase the radiation gain by almost 3 dB. The AMC-grounded antenna features a more compact profile than the PEC-grounded one.

\bibliographystyle{IEEEtran}
\bibliography{IEEEexample}

\end{document}